# On the existence of Einstein-Podolsky-Rosen Channels

B. C. Sanctuary
Department of Chemistry,
McGill University
801 Sherbrooke Street W
Montreal, PQ, H3A 2K6, Canada

**Abstract**

A physical theory without interpretation is mathematics. Since there are no paradoxes in science, only incorrect interpretations of phenomena or inadequate theories, it is necessary to use a consistent interpretation of quantum mechanics that makes physical sense and satisfies the experimental facts. Quantum "teleportation" provides such an example because the current treatments rely on unexplained connections between separated correlated particles. In this comment, a mechanism is suggested that avoids the instantaneous wave function collapse between non-interacting entangled particles at space like separations. This mechanism requires a statistical ensemble interpretation of the wave function.

**Comment**

In 1993, Bennett[1] *et al* published a seminal paper titled, *Teleporting an Unknown State via Dual Classical and Einstein-Podolsky-Rosen Channels*. This comment does not find fault with the mathematics of that paper but rather questions the existence of EPR channels and the viability of the process called quantum teleportation.

It is often possible to come up with a logical model that describes a phenomenon or satisfies certain observations. Devoid of interpretation, however, such theories are simply mathematics. For example, Ptolemy's theory reproduces the trajectories of the planets yet has nothing to do with gravity. Interpretation of the wave[2] function is therefore essential to validate the mathematical formalism of quantum theory.

Quantum "teleportation" depends on instantaneous wave function[3] collapse over space-like separations when an entangled EPR pair is beyond the range of any interaction. These EPR correlations provide a means of sharing the information present in an entangled pair. The mechanism by which the connection results in the sharing of information has not been satisfactorily established. Words like quantum "weirdness", "magic" and "spookiness" are commonly used to describe the type of non-local phenomena associated with quantum teleportation. More clearly Maudlin[4] has characterized this connection between space-like separated entangled particles as being

- Unattenuated, meaning independent of distance between the two EPR particles.
- Discriminating, meaning that it operates only between specific particles (e.g. EPR pairs),
- Instantaneous, meaning faster than the speed of light.

In this comment, the concept that the wave function describes the state of a pair of particles is replaced with the wave function describing a statistical ensemble of possible EPR pairs[5]. The phenomenon of quantum "teleportation" then becomes a more prosaic "state selection" process. That is Alice's unknown photon state can only form a Bell state with one of a pair of entangled photons, call it photon 2, if that photon has the same characteristics as Alice's photon 1. Only a small sub-ensemble of all the possible states originating from the random source, which produce entangled photons 2 and 3, can have the required property, so Alice's photon 1 and Bob's photon 3 have, up to a known transformation, the same attributes.

To be specific, an entangled singlet state is one of the four Bell states,

$$|\Psi^-_{12}\rangle = \frac{1}{\sqrt{2}}\left[|+\rangle^1_{\hat{z}}|-\rangle^2_{\hat{z}} - |-\rangle^1_{\hat{z}}|+\rangle^2_{\hat{z}}\right]. \tag{1}$$

Such a state is common in quantum theory with the two outcomes for each spin, "+" and "-", relative to a quantization axis that is arbitrarily chosen to be in the $z$ direction. The singlet state has total angular momentum of zero and is completely isotropic in spite of the quantization axis, $\hat{z}$.

The conventional interpretation asserts that when Alice's photon forms a new entangled pair with photon 2, Bob's photon collapses into one of four possible states, one of which is identical to Alice's original photon state and three others related to her state by a unitary transformation. Figure 1 illustrates these steps. To complete the process, so Bob can distinguish which of the four possible states actually matches Alice's photon state, Alice must communicate to Bob over classical channels. This last step is not shown in Figure 1.

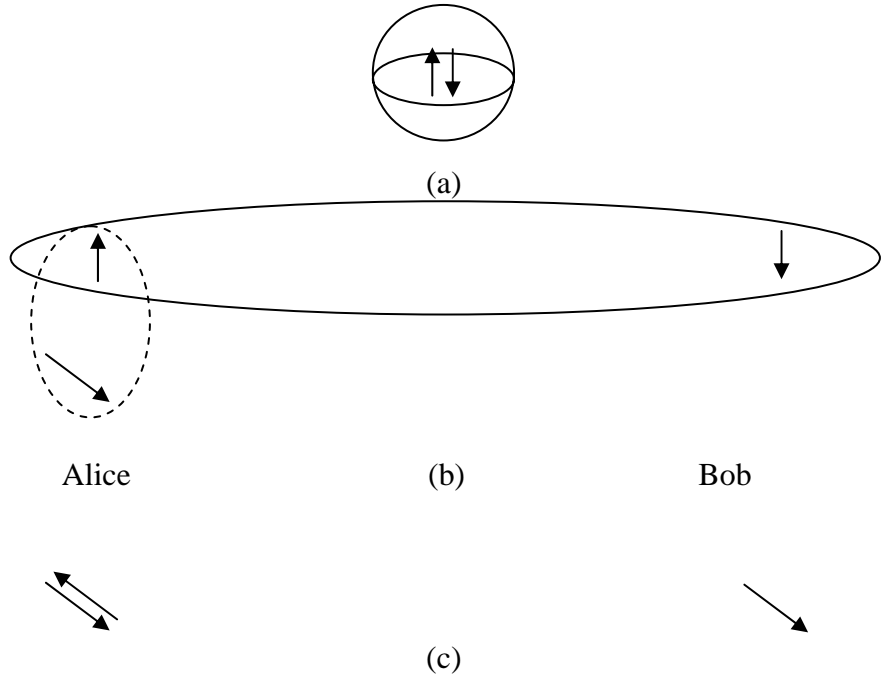

Alice (b) Bob

(c)

Figure 1. (a) Depiction of an isotropic singlet state as a sphere. The EPR pair is indicated by the up and down arrows but the orientation is arbitrary since the wave function can collapse into a state with quantization axis in any direction in 3D real space. (b) The singlet pair has separated, one photon moving towards Alice and the other towards Bob. They remain entangled over this distance. Alice's photon can interfere with the left-moving photon to form a new entangled pair. (c) The singlet state collapses and forms a Bell state with Alice's photon. Instantaneously, the photon arriving at Bob's location collapses over space-like distances into the state determined by the polarization of Alice's photon. The process is non-local.

The instantaneous collapse of the wave function over large distances is considered to be experimental evidence for quantum non-locality. By this it is meant that such a situation as encountered in "teleportation" cannot be explained by a local hidden variable theory. The results require more than knowledge alone of the local settings of the photon detectors and the properties of the photons at each location. In other words a non-local hidden variable theory is required and such theories violate Bell's inequalities[6]. It is well known that the singlet state can lead to violation of Bell's inequalities[7].

If one considers that the wave function describes a statistical ensemble of similarly prepared states, then there is no need for wave function collapse over space-like separations. It was Einstein who first articulated the statistical interpretation of the wave function at the 1927 Solvay conference. Adapting this concept to the present case means generalizing Eq.(1) to

$$\left|\Psi_{12}^{-}\right\rangle = \frac{1}{\sqrt{2}}\left[\left|+\right\rangle_{\hat{\mathbf{P}}}^{1}\left|-\right\rangle_{\hat{\mathbf{P}}}^{2} - \left|-\right\rangle_{\hat{\mathbf{P}}}^{1}\left|+\right\rangle_{\hat{\mathbf{P}}}^{2}\right] . \tag{2}$$

where the axis of quantization is changed from $\hat{z}$ to $\hat{\mathbf{P}}$. This does not change the fact the singlet state is still isotropic and makes no difference to the original theory[1]. In this case, the wave function can form from any $\hat{\mathbf{P}}$. Every possible value of the unit vector $\hat{\mathbf{P}}$ leads to the same indistinguishable isotropic singlet state. Note however that the axis of quantization must be the same for both spins of the entangled pair in Eq.(2) Since $\hat{\mathbf{P}}$ is not explicitly evident on the LHS of Eq.(2), it is a quantum hidden variable. However the technique of quantum "teleportation" makes it possible to reveal the particular hidden variable since it must match that of Alice's photon.

Consider that Alice's photon, the one whose properties we wish to teleport, is in a state defined by a quantization axis $\hat{\mathbf{P}}'$. When her photon encounters one of the photons from the entangled photon pair, in order to form a new entangled state, only one component of the ensemble is suitable. This component must have $\hat{\mathbf{P}} = \hat{\mathbf{P}}'$. No other component can produce a singlet pair with Alice's photon as required by the theory. Therefore the photon that heads to Bob, which originated from the same initial entangled photon pair that Alice's encountered, must also have $\hat{\mathbf{P}} = \hat{\mathbf{P}}'$. In this way, the properties of Alice's photon are the same as Bob's up to a transformation, which is dictated by the calculation and the actual Bell state measured at Alice's location. Figure 2 depicts this mechanism.

Quantum "teleportation" experiments have extremely low detection rates. This is consistent with the statistical ensemble interpretation where only one axis of quantization, from the ensemble of possibilities, can be selected as to form the required Bell state with Alice's photon.

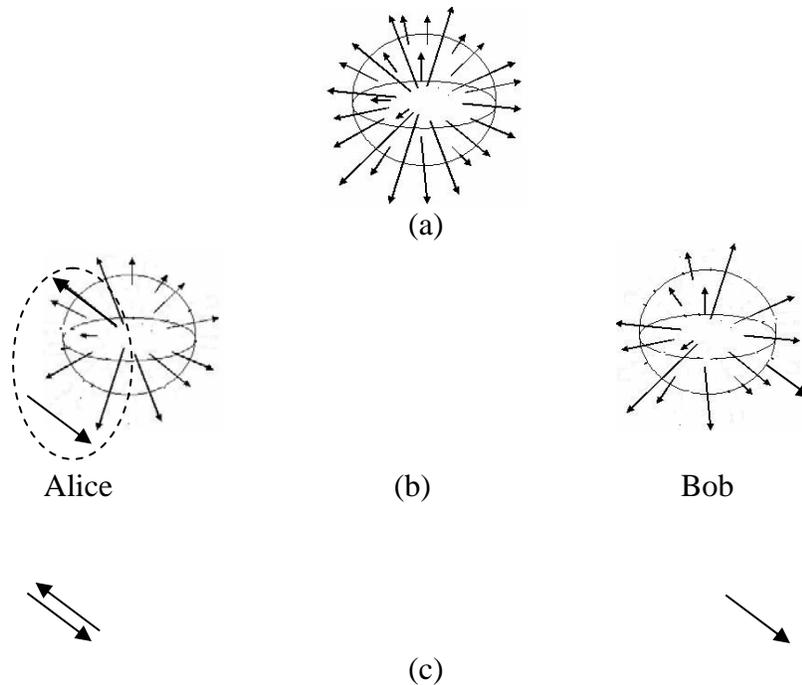

Figure 2. (a) A depiction of a statistical ensemble of EPR pairs that form a singlet state. Even though the axes of quantization are hidden, some are shown for clarity. The state remains isotropic, but any axis of quantization leads to an isotropic singlet state. (b) The EPR pairs moves apart and remains entangled. For every photon moving to Alice in one state there is a photon moving to Bob in the opposite state with respect to the same axis of quantization. The component that can form a Bell state with Alice's photon is highlighted as is its EPR partner at Bob's location. (c) Alice's photon can only form a Bell state with an EPR photon that has the same axis of quantization as her polarization state (see Eq.(2)). Therefore the photon at Bob's location must have the same quantization axis as Alice's. There is no wave function collapse and the effect is local and real.

By asserting the statistical ensemble interpretation of the wave function, quantum "teleportation" becomes a local phenomenon. No non-quantum hidden variables are needed since the ensemble of quantization axes provides the appropriate elements of reality. No long range instantaneous collapse of the wave function is necessary since the correlation is created at the source by virtue of the EPR pair sharing a common axis of quantization. Finally violation of Bell's inequalities indicates, in this instance, that the photons are entangled, and hence the effect depends, in part, on the quantum interference terms present in the entangled pair.

**Acknowledgement:**

*This work was supported by a research grant from the Natural Sciences and Research Council of Canada (NSERC).*